# On the Electride Nature of Na-hP4

Stefano Racioppi,*[a] Christian V. Storm,[b] Malcolm I. McMahon,[b] Eva Zurek*[a]

In memoriam of Neil Ashcroft, who died in March 2021.

[a]  S. Racioppi, Prof. E. Zurek
    Department of Chemistry
    State University of New York at Buffalo
    777 Natural Science Complex
    E-mail: sraciopp@buffalo.edu
    E-mail: ezurek@buffalo.edu
[b]  Christian V. Storm, Prof. Malcolm I. McMahon
    SUPA, School of Physics and Astronomy, and Center for Science at Extreme Conditions
    The University of Edinburgh
    Peter Guthrie Tait Road, Edinburgh EH9 3FD, United Kingdom

**Abstract**
Early quantum mechanical models suggested that pressure drives solids towards free-electron metal behavior where the ions are locked into simple close-packed structures. The prediction and subsequent discovery of high-pressure electrides (HPEs), compounds assuming open structures where the valence electrons are localized in interstitial voids, required a paradigm shift. Our quantum chemical calculations on the iconic insulating Na-hP4 HPE show that increasing density causes a 3s → 3pd electronic transition due to Coulomb repulsion between the 1s2s and 3s states, and orthogonality of the 3pd states to the core. The large lobes of the resulting Na-pd hybrid orbitals point towards the center of an 11-membered penta-capped trigonal prism and overlap constructively, forming multi-centered bonds, which are responsible for the emergence of the interstitial charge localization in Na-hP4. These multi-centered bonds facilitate the increased density of this phase, which is key for its stabilization under pressure.

In their classic paper *On the Constitution of Sodium at Higher Densities*, Neaton and Ashcroft[1] showed that the nearly-free-electron model of Wigner and Seitz[2] was valid only at ambient or near-ambient pressures. At higher densities the effects of Pauli exclusion and orthogonality, coupled with a 3s → 3pd transition (Figure 1), stabilized low-coordinated structures whose electron density was localized in interstitial regions, heralding the onset of a metal-to-insulator transition. Almost a full decade later this intuitive prediction was verified by experiments that characterized the Na-hP4 phase.[3] First principles calculations yielded a band gap of ~1.3 eV at 200 GPa for this newly discovered allotrope, whose electronic structure was reminiscent of an electride – a compound where electrons localized at interstitial sites assume the roles of anions, and, in sodium, the ionic cores play the role of cations. Since then, many experimental and theoretical studies have tried to answer the question: "Will every element turn into an electride when squeezed?"[4,5] Herein, detailed quantum chemical calculations are performed on Na-hP4 as a model for a HPE, addressing long-standing questions regarding the emergence of the electride state under pressure.

Chemists have known of electrides since Humphry Davy's experiments on alkali metals in ammonia solutions in the 19$^{th}$ century.[6] The seminal work of Dye in the 1980s inspired rigorous studies of molecular electrides,[7,8] culminating in the X-ray assisted visualization of the residual electron density of an organometallic electride.[9] Examples of ambient-pressure inorganic crystalline electrides include [$Ca_{24}Al_{28}O_{68}$]$^{4+}$[4e$^-$][10] and several layered materials,[11] some possessing industrially-useful catalytic properties.[12] Theoretical calculations predict that in addition to alkali and alkaline earth metals,[13,14] aluminum[4] and carbon,[5] even exotic compounds like $Na_2He$[15] can become electrides when squeezed, suggesting that at high pressure the electride state may actually be quite common.

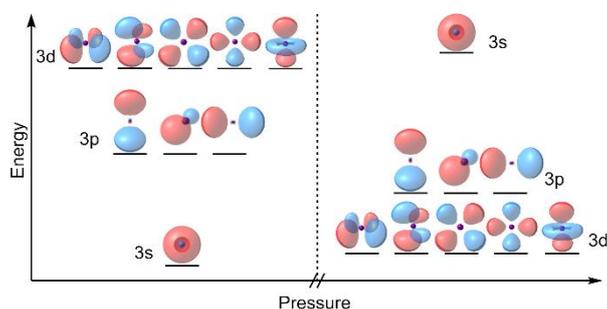

**Figure 1.** Schematic illustrating the concept of energy-level reorganization upon compression for an isolated atom of sodium, based on the expected Coulomb repulsion exerted by the core, resulting in 3s → 3p and 3p → 3d electronic transitions. Though interatomic interactions in the solid state may affect the orbital ordering, it will still differ significantly from the orbital ordering found at ambient conditions.

Theoretically, electrides can be characterized using several electron density (ρ) based methods,[16,17] such as the Quantum theory of Atoms in Molecules (QTAIM), as well as Electron Localization Functions (ELFs) and Non-Covalent Interaction analyses. As suggested from studies on molecular systems,[17] an electride should fulfill at least three topological criteria and possess: 1) a non-nuclear attractor (NNA, local maxima of the electron density that does not coincide with an atomic position), 2) negative value of $\nabla^2\rho$ at the NNA position (meaning that charge accumulates at that position), and 3) the presence of an ELF basin. The NNA lends some kind of identity to the electride, with an unambiguous physics definition for its volume and charge.[18] Even though this characterization appears to be quite straightforward, there is, to date, no theory that can unequivocally explain and predict HPE formation. The most thorough attempt thus far is the theory of Interstitial Quasi-atoms (ISQs) by Miao and Hoffmann.[19,20] If the energies of orbitals centered on interstitial sites devoid of an atomic nucleus, which possess s orbital character, become lower than the energies of atom-centered orbitals upon densification, ISQ theory predicts that electrons will move off the atoms to the interstitial regions forming a HPE. Calculations on lithium and sodium atoms confined in a pressure-simulating helium matrix corroborate this theory.[19]

Though ISQ-theory is chemically appealing it does not explain HPE formation in all elements, and the original calculations, which were carried out over a decade ago, at times lack predictive power. For example, it has long been hypothesized that elements such as Ca[21] and Cs[22,23] become electrides because their valence electrons undergo pressure induced s → d transitions. Moreover, though ISQ theory predicts that Tl becomes a HPE[19], calculations on experimentally observed structures suggest that pressure induced sp→d transitions, not considered by Miao and Hoffmann for this element, inhibit the formation of the electride state.[24] Finally, though some ambient pressure phases of pure metals, including Sc and Al together with alkali and alkaline earth metals,[25,26] and even gas-phase clusters of sodium and lithium[18,27,28] possess charge localization loci (corresponding to NNAs) and all the topological requirements to be classified as electrides, they would not necessarily be predicted as such by ISQ-theory. In fact, charge concentrations in interstitial regions in electropositive metals are traditionally attributed to the formation of metallic bonds, and not to electride formation.[25,26] Therefore, it seems that further investigations are still necessary to understand the forces driving the formation of HPEs.

Na-hP4 is one of the most extensively characterized HPEs,[3,14,29,30] being simple in structure, chemical composition, and relatively easy to form upon compression (by ~200 GPa in a diamond anvil cell). Surprisingly, there is no general consensus as to why Na-hP4 becomes an electride. In addition to the aforementioned combined effects of Pauli exclusion, Coulomb repulsion and orthogonality,[1] and the ISQ-theory,[19,31] the proposed mechanisms include: repulsions due to core exclusion,[32] and valence s-p[29] or p-d hybridizations.[1,3,14,30] Herein, we present the results of careful quantum chemical computations designed to answer the following questions for Na-hP4. i) What are the orbitals responsible for the formation of the electride state? ii) What is the role of the core-electrons? And iii) Are the localized electrons best described as those centered on ISQs, or do they still belong to the surrounding atoms, being manifestations of the build-up of electron density arising from multi-centered chemical bonds?

To answer these questions, density functional theory calculations were performed using the HSE06 functional, coupled with either planewave or atom centered basis-sets on the experimental Na-hP4 geometry at select pressures[3] (Section S1). Projection of the electronic band structure onto localized orbitals is the most obvious approach to answer the first question. However, care must be taken in performing such an analysis to avoid inaccurate projections (Figure S1-S3). The planewave based codes used by us and others, associate to each atom a spherical volume that encloses the projection onto fixed atomic orbitals, and therefore they only have a qualitative meaning. Using the default parameters provided in one of the most popular band structure programs, VASP, such calculations generally suggest that the valence states of Na-hP4 exhibit primarily Na s[20,31] or hybrid s-p character.[29] However, these results are sensitive to the size of the projection's sphere. Unless the spheres perfectly cover the entire space of the unit cell without overlapping with each other - an impossible task in Na-hP4 due to the presence of the cavities - the projection cannot be complete (Section S2). Optimization of the sphere size based on a Bader analysis, or by choosing spheres whose volumetric sum is the same as the total volume of the cell suggests various degrees of s-p-d hybridization instead (Figure S2).

When localized basis functions are employed, the electronic states are projected on the very same orbitals used to generate the band structure. Moreover, it becomes possible to add or remove orbitals from the basis, enabling studies that consider how the presence or absence of specific functions affects the electronic structure of a compound. Therefore, we constructed an optimized basis (named *N*a) containing all core states, as well as sodium 3spd functions that yielded the density of states, band gap, ELF, topology of the electron density, and NNA charges that agreed with the results of planewave codes (Figure 2, Table 1). Removing the 3d (*Na-d*) or 3pd (*Na-pd*) functions from the basis restricts Na to adopt either a 3sp or 3s electronic configuration, predicting metallic behavior for Na-hP4, sided by the disappearance of the charge concentrations in the cavities. This is shown most clearly when the ELF obtained from the *Na-d* basis is subtracted from that obtained with the optimized basis (Figure 2b). Adding a single s (*Na-pd+1s*) function, or three s functions with different orbital exponents (*Na-pd+3s*) in the interstitial regions treats Na as having a $3s^1$ configuration, but it allows the possibility for electrons to be transferred to ISQ-like orbitals. These basis sets yielded band gaps that were too large (2.5-2.9 eV), and ELFs that were too localized (*e.g.*, Figure 2c). Adding back the p functions to both basis-sets (*Na-d+1s* or *Na-d+3s*), yielded band gaps that were somewhat smaller than expected (0.8-1.2 eV), but not unreasonable, and ELF plots that even resembled the planewaves results (Figure S4).

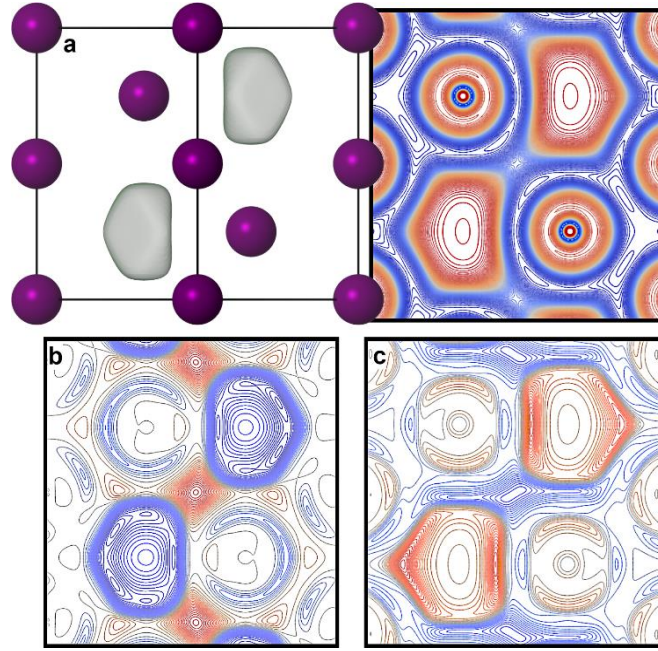

**Figure 2.** a) 3-D ELF basin (isovalue = 0.8) and 2-D plots along the (110) plane of Na-hP4 at 190 GPa (*Na* basis set). 2-D ELF difference calculated between the b) [*Na-d* – *Na*]; c) [*Na-pd+1s* – *Na*] basis sets. The color scale in the 2-D plot of (a) goes from 0 (blue) to 1 (red), while in (b) and (c) it goes from -0.25 (blue) to +0.25 (red).

ELF provides a visual and qualitative analysis of the electronic structure. Could a quantitative measure be used instead? A feature of the electron density characteristic only of electrides[17] and some pure metallic bonds,[25] is the presence of NNAs at the center of the ELF's lobes. Therefore, to unambiguously distinguish between the atom-centered basis-sets that could potentially describe the electronic structure of Na-hp4 (*Na*, *Na-d+1s* and *Na-d+3s*) we analyzed the topological features of the NNAs, *i.e.*, electron density ($\rho$) and Laplacian of the electron density ($\nabla^2\rho$), by means of QTAIM. As show in Table 1 and Figure S4, an electron density topology that agreed with the planewave results could only be achieved when the electrons in the cavities were generated via sodium pd hybrids as in the optimized *Na* basis. If ISQ centered s-orbitals were employed instead, the electron densities in the cavities were too localized (with large absolute values of $\rho_{NNA}$ and $\nabla^2\rho_{NNA}$).

**Table 1.** Band gaps [eV], values of electron density ($\rho_{NNA}$) [e/bohr$^3$] and its Laplacian ($\nabla^2\rho_{NNA}$) [e/bohr$^5$], integrated charge ($q_{NNA}$) [e] at the NNA positions calculated with atomic-orbital (AO) and plane-wave (PW) basis sets (HSE06 functional). The number of valence electrons used explicitly in the plane-wave calculations is provided.

| Basis-set (AO) | Band-Gap | $\rho_{NNA}$ | $\nabla^2\rho_{NNA}$ | $q_{NNA}$ |
|---|---|---|---|---|
| *Na* | 1.25 | 0.048 | -0.035 | -1.10 |
| *Na-d* | 0.00 (metallic) | N/A | N/A | N/A |
| *Na-d+1s* | 0.78 | 0.061 | -0.100 | -1.14 |
| *Na-d+3s* | 1.19 | 0.057 | -0.763 | -1.14 |
| *Na-pd* | 0.00 (metallic) | N/A | N/A | N/A |
| *Na-pd+1s* | 2.50 | 0.077 | -0.159 | -1.20 |
| *Na-pd+3s* | 2.92 | 0.062 | -0.862 | -1.15 |
| **Valence Electrons (PW)** | **Band-Gap** | $\rho_{NNA}$ | $\nabla^2\rho_{NNA}$ | $q_{NNA}$ |
| 9 | 1.23 | 0.049 | -0.043 | -1.08 |
| 7 | 1.24 | 0.049 | -0.042 | -1.07 |
| 1 | 1.25 | 0.051 | -0.047 | -1.02 |

Therefore, the most appropriate way to view the valence electronic states of Na-hp4 is as 3pd hybrids, and any extra functions at the ISQ sites need not be invoked, in-line with Neaton and Ashcroft's original thesis. Upon densification, the Na 3s orbitals experience repulsion from the core, but the 3p and 3d do not because of orthogonality. This effect, quantified through an analysis of the intra-atomic Crystal Orbital Overlap / Hamilton Populations (Table S3) is responsible for the pd hybridization, but it does not imply that the valence pd-electrons are expelled from the atom into the ISQ sites. To test whether the 2s and 2p orbitals affect the electronic structure of Na-hP4, either by polarizing or by repelling the outermost electron, additional planewave calculations were performed explicitly treating either 7 or 1 valence electron. The band gaps and topological descriptors obtained did not depend upon the number of valence electrons considered (Tables 1, S1-S2). Finally, our QTAIM analysis (Table 1) located NNAs inside the cavities of Na-hP4, characterized by negative Laplacians and ELF lobes, so the aforementioned electride-state-checklist is fulfilled.[17]

To better understand how such Na pd hybrids yield the electron density calculated for Na-hp4, we constructed a Wannier-like function (WF)[33] that exactly reproduced the doubly occupied valence band of this allotrope (Figure 3a). A constructive overlap of pd hybrids placed on a square lattice yields an $A_{1g}$ molecular orbital with a build-up of electron density in the interstitial (Figure 3b/3c). Analogously, the NNA-centered WF in Na-hP4, which clearly resembles the ELF feature associated with the localized electron, results from an "in-phase" overlap of pd hybrids on 11 Na atoms. This WF is therefore a consequence of a multi-centered bond formed within a penta-capped trigonal prism whose nearest neighbor Na-Na distances vary from 1.995 to 2.920 Å at 190 GPa, with the distance to the center of the cavity ranging from 1.686 to 2.135 Å. The concentration of electron density in the cavities allows Na atoms to be closer to each other resulting in a decreased volume, which is the reason why Na-hP4 becomes more stable than bcc under pressure (Section S6). The integrated Bader charge within the NNA, approx. 1e, resembles calculations performed for analogous systems,[27] and the rest of the electron density, shared within the framework, accounts for the smaller lobes of the hybrid orbitals. A picture where the charge density within the cavity results from multi-centered bond formation, and the rest is distributed around the Na atoms in the framework, is a less ionic situation then a doubly occupied localized ISQ, and more in line with the semi-conducting band-gap of ~1.3 eV (Table 1).

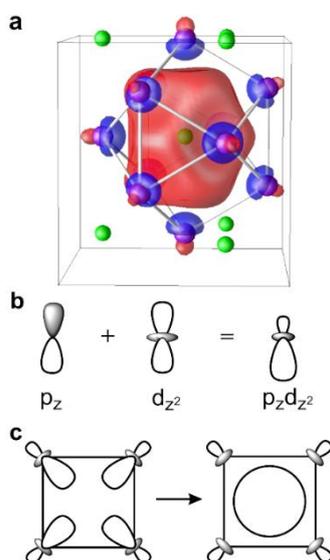

**Figure 3.** a) The NNA-centered Wannier-like function that describes the valence band in Na-hP4 at 320 GPa. Na atoms /NNAs are purple/green. Schematic of b) $p_z$ and $d_{z2}$ orbital hybridization; c) a multicentered bond formed from pd hybrids pointing towards the square's interior, resulting in a cavity-centered molecular orbital.

Interestingly, QTAIM identifies the localized electrons within this HPE as well-defined topological entities connected with the Na atoms (Figure S6-S7, Table S4). Although "topologically connected" does not necessarily mean "chemically bonded", the presence of NNAs, and their contacts with the surrounding atoms, is an intrinsic characteristic of several metallic compounds, naturally emerging upon formation of chemical bonds.[26] In fact, NNAs, and the familiar ELF blobs associated with the electride state, are not a consequence of pressure either, but they exist in gas-phase,[18,26,27] as well as in solid state compounds at ambient conditions.[25] Moreover, the presence of NNAs in cavities, like in Na-hP4, instead of in-between atoms, is again a consequence of the electronic transition to (p)d-orbitals, which provide the extra directionality off the interatomic axes (Section S7). In fact, removing the (p)d-orbitals from Na (Figure 2, S4b), led the charge concentrations in the cavities to disappear. Therefore, we conclude that electrides like Na-hP4, and many others involving electropositive elements[9,25] can be described by means of (multicentered) chemical bonding schemes, which might become ubiquitous under pressure.[34]

Thus, we have obtained the following answers to the questions posed above: i) atom-centered pd-hybrid orbitals whose overlap leads to charge concentrations in the cavities define the electride state in Na-hP4; ii) the only effect of sodium's core upon densification is to hasten the electronic 3s→3pd transition, but it plays no role in the generation of the electride state associable with "squeezing out" electrons from the atom; iii) the electrons in the cavities of Na-hP4 are not unbound, isolated objects, but they are generated by the sharing of the electron density of the surrounding sodium atoms, forming exceptional multicenter-bonds.

By considering all of the theories proposed for the formation of the electride state in high-pressure systems,[1,3,19,29,32] we illustrated conclusively that the arguments of Neaton and Ashcroft are the only valid quantum mechanical explanation for the emergence of this state in compressed Na, which behaves similar to K, Rb and Cs under pressure. However, our work goes beyond the physics picture painted by Neaton and Ashcroft, connecting it with chemical concepts of bonding. Specifically, we show that a Wannier-like function that spans the valence bands of Na-hP4 can be interpreted as pd hybrid orbitals that overlap in a constructive way in the interstitial region where the excess electron is found, as in a multi-centered chemical bond. The sharing of electrons inside a cavity within an Na$_{11}$ cluster, instead of along a nearest-neighbor contact as in a two-center chemical bond, is in fact a consequence of the hybridization between orbitals of higher angular momentum, here p and d, that confer extra dimensionality to the interatomic interactions. The analysis presented herein is the first step towards a predictive theory for the discovery of solid-state electrides based upon crystal structure prediction methods and QTAIM descriptors.

## Acknowledgements

Funding for this research is provided by the *Center for Matter at Atomic Pressures* (CMAP), a National Science Foundation (NSF) Physics Frontier Center, under Award PHY-2020249, and by Grants No. EP/R02927X/1 and EP/S022155/1 from the UK Engineering and Physical Sciences Research Council (EPSRC). Calculations were performed at the Center for Computational Research at SUNY Buffalo (http://hdl.handle.net/.10477/79221). Partial funding for this research is provided by the U.S. Department of Energy, Office of Science, Fusion Energy Sciences funding the award entitled *High Energy Density Quantum Matter*, under Award No. DE-SC0020340.

**Keywords:** Electride • High Pressure • Sodium • Alkali metals • Chemical Bonding

**Table of Contents**

**1.0 Computational Details**
The goal of the present work is to attain a microscopic understanding of the electronic origin of the electride state in Na-hP4. Towards this end, periodic Density Functional Theory (DFT) calculations using both an atomic orbital based code, Crystal17,[1] and a plane-wave based code, VASP (version 5.4.1)[2], were performed. Though the electronic structure of this insulating high-pressure phase has been interrogated in the past, the analyses were performed on the basis of results obtained with density functional theory programs that employ plane-wave basis sets.[3–6] The conclusions reached in these studies regarding the nature of the electride state differ, motivating our use of both plane-wave and atom-centered basis sets. To reduce the potential discrepancies that could be obtained using the two different codes, we opted to employ the experimental hP4 geometry measured by Ma *et al.*[3] at 190 GPa (lattice parameters a = b = 2.92 Å, c = 4.27 Å) and at 320 GPa (lattice parameters a = b = 2.784 Å, c = 3.873 Å) in our calculations.

**1.1 Crystal17 Calculation Details**
The HSE06[7–9] exchange-correlation functional was used for the presented results. The k-point grid was generated with the Monkhorst-Pack method, using a shrinking factor of 32 along the reciprocal lattice vectors (32x32x32 grid in k-space), while the convergence threshold of the total energy was set to $10^{-8}$ Ha.

One of the main advantages of using atom centered basis functions is the possibility of performing all-electron calculations, meaning that the core electrons are treated explicitly. However, the typical drawback of this approach is the limitation in the number of basis functions that can be associated to each atom. Moreover, most of the available basis sets are optimized for specific systems and at ambient pressure. For these reasons, we have optimized some of the exponents associated with the Gaussian functions of the largest basis set for which we could achieve convergency in the self-consistent calculations, finding the combination that minimized the energy of Na-hP4 at the experimental geometry (see below the reported basis-set, which we refer to as the *Na* basis) and yielded a densities of states (DOS) plot that matched well with the VASP calculations.

To study the effect of the d-orbitals on the generation of the electride state, as well as the presence of basis functions explicitly placed on the empty regions where the interstitial electron density is found (as a model for the interstitial quasi-atom or ISQ), we tested a series of basis-sets: 1) the basis-set optimized for Na-hP4, which includes Na 3p- and 3d-orbitals in addition to the 3s orbital, named *Na*; 2) the optimized basis-set, excluding the Na 3d-orbitals only, named *Na-d*; 3) the optimized basis-set, excluding the Na 3d-orbitals, but including an s-orbital at the ISQ position, named *Na-d+1s*; and 4) the optimized basis-set, excluding the Na 3d-orbitals, but including three s-orbitals (with different orbital exponents) at each ISQ position, named *Na-d+3s*; 5) the optimized basis-set, excluding the Na 3p3d-orbitals, named *Na-pd*; 6) the optimized basis-set, excluding the Na 3p3d-orbitals, but including an s-orbital at the ISQ position, named *Na-pd+1s*; and 7) the optimized basis-set, excluding the Na 3p3d-orbitals, but including three s-orbitals (with different orbital exponents) at each ISQ position, named *Na-pd+3s*. The *Na-pd+1s* and *Na-pd+3s* models will force Na to retain a $3s^1$ valence configuration, but allow the electrons to hop into the ISQ sites. On the other hand, *Na-d+1s* and *Na-d+3s* allows for both a 3p polarization on Na and the hopping of the valence electron into the ISQ sites. We uploaded the modified basis-set in Zenodo (ID: 10.5281/zenodo.8369654).
Atomic *Na* basis set (Crystal17 format).

```
11  13
0 0 10 2.0 1.0
```

```
    379852.2008100         0.20671384468D-04
     56886.0063780         0.16070466617D-03
     12942.7018380         0.84462905848D-03
      3664.3017904         0.35519026029D-02
      1194.7417499         0.12754034468D-01
       430.98192917        0.39895462742D-01
       167.83169424        0.10720154498
        69.306669040       0.23339516913
        29.951170886       0.36333077287
        13.380791097       0.30544770974
0 0 3 2.0 1.0
       121.74011283        0.36142427284D-01
        37.044143387       0.28820961687
        13.995422624       0.79337384869
0 0 1 0.0 1.0
         3.4427            1.0000000
0 0 1 0.0 1.0
         1.8293            1.0000000
0 0 1 0.0 1.0
         0.781755          1.0000000
0 0 1 0.0 1.0
         0.20              1.0000000
0 2 8 6.0 1.0
       690.77627017        0.37478518415D-03
       163.82806121        0.31775441030D-02
        52.876460769       0.16333581338D-01
        19.812270493       0.59754902585D-01
         8.1320378784      0.15879328812
         3.4969068377      0.29049363260
         1.5117244146      0.36368131139
         0.64479294912     0.28195867334
0 2 1 0.0 1.0
         2.14048           1.0000000
0 2 1 0.0 1.0
         1.11195           1.0000000
0 2 1 1.0 1.0
         0.20              1.0000000
0 3 1 0.0 1.0
         1.99145           1.0000000
0 3 1 0.0 1.0
         0.721653          1.0000000
0 3 1 0.0 1.0
         0.23              1.0000000
```

The *Na-d* and *Na-pd* basis sets were then generated by removing the 3d and the 3p3d basis functions from the *Na* basis set.

Atomic **Na-d** basis set (Crystal17 format).

```
11  10
0 0 10 2.0 1.0
    379852.2008100         0.20671384468D-04
```

```
     56886.0063780          0.16070466617D-03
     12942.7018380          0.84462905848D-03
      3664.3017904          0.35519026029D-02
      1194.7417499          0.12754034468D-01
       430.98192917         0.39895462742D-01
       167.83169424         0.10720154498
        69.306669040        0.23339516913
        29.951170886        0.36333077287
        13.380791097        0.30544770974
0 0 3 2.0 1.0
       121.74011283         0.36142427284D-01
        37.044143387        0.28820961687
        13.995422624        0.79337384869
0 0 1 1.0 1.0
         3.4427             1.0000000
0 0 1 0.0 1.0
         1.8293             1.0000000
0 0 1 0.0 1.0
         0.781755           1.0000000
0 0 1 0.0 1.0
         0.20               1.0000000
0 2 8 6.0 1.0
       690.77627017         0.37478518415D-03
       163.82806121         0.31775441030D-02
        52.876460769        0.16333581338D-01
        19.812270493        0.59754902585D-01
         8.1320378784       0.15879328812
         3.4969068377       0.29049363260
         1.5117244146       0.36368131139
         0.64479294912      0.28195867334
0 2 1 0.0 1.0
         2.14048            1.0000000
0 2 1 0.0 1.0
         1.11195            1.0000000
0 2 1 0.0 1.0
         0.20               1.0000000
```

Atomic **Na-pd** basis set (Crystal17 format).

```
11  7
0 0 10 2.0 1.0
    379852.2008100          0.20671384468D-04
     56886.0063780          0.16070466617D-03
     12942.7018380          0.84462905848D-03
      3664.3017904          0.35519026029D-02
      1194.7417499          0.12754034468D-01
       430.98192917         0.39895462742D-01
```

```
   167.83169424       0.10720154498
    69.306669040      0.23339516913
    29.951170886      0.36333077287
    13.380791097      0.30544770974
0 0 3 2.0 1.0
   121.74011283       0.36142427284D-01
    37.044143387      0.28820961687
    13.995422624      0.79337384869
0 0 1 1.0 1.0
     3.4427           1.0000000
0 0 1 0.0 1.0
     1.8293           1.0000000
0 0 1 0.0 1.0
     0.781755         1.0000000
0 0 1 0.0 1.0
     0.20             1.0000000
0 2 8 6.0 1.0
   690.77627017       0.37478518415D-03
   163.82806121       0.31775441030D-02
    52.876460769      0.16333581338D-01
    19.812270493      0.59754902585D-01
     8.1320378784     0.15879328812
     3.4969068377     0.29049363260
     1.5117244146     0.36368131139
     0.64479294912    0.28195867334
```

Regarding the *Na-d+1s*, *Na-d+3s, Na-pd+1s* and *Na-pd+3s* basis-sets, the following basis functions were added to *Na-d* and *Na-pd* at the ISQ positions using dummy-atoms:

***+1s)***
```
0 1
0 0 1 0 1
  0.1795111000     1.00000000
```

***+3s)***
```
0 3
0 0 3 0.0 1.0
  34.061341000    0.00602519780
   5.1235746000   0.04502109400
   1.1646626000   0.20189726000
0 0 1 0.0 1.0
   0.5157455100   1.00000000000
0 0 1 0.0 1.0
   0.1795111000   1.00000000000
```

**1.2 VASP Calculation Details**

The electronic structure was calculated using the Vienna *Ab Initio* Simulation Package (VASP), version 5.4.1,[2,10] with the HSE06[7–9] exchange-correlation functional. The projected augmented wave (PAW) method[11] was used to treat the core states in combination with a plane-wave basis set with an energy cut-off of 910 eV for the HSE06 functional, as proposed by Ma *et al.*,[3] (and dictated by convergency). To test the effect of core-electrons, we used PAW-PBE Na POTCARs where 1, 7 and 9 valence electrons were treated explicitly. In addition, the projected density of states (pDOS) were also calculated with the PBE[12,13] exchange-correlation functional with an energy cut-off of 2000 eV.

The k-point grids were generated using the Γ-centred Monkhorst−Pack scheme and the number of divisions along each reciprocal lattice vector was chosen so that its product with the real lattice constant was equal to 40 Å.

## 1.3 Details on the Topological and Orbital Analyses

The topological analysis of the electron density, based on the Quantum Theory of Atoms in Molecules (QTAIM),[14] was performed using the TOPOND package implemented in Crystal17,[15] and the Critic2[16] code for the calculations performed with VASP. The Crystal Orbital Overlap Population (COOP) and Crystal Orbital Hamilton Population (COHP) analyses[17] were performed with Crystal17.[18]

## 2.0 Projected-orbital Analysis

There are some important technical differences to be addressed when the projection of the density of states (pDOS) is performed either with an atomic orbital-based code, like Crystal17, or with a plane-wave code, like VASP. In Crystal17 the projection is performed on the exact same basis-set used in the calculation, whereas in VASP, the projection onto standardized localized orbitals is performed for each band enclosed in the Wigner-Seitz atomic volume generated by a fixed radius. Clearly, in the latter case the partition process in not unambiguos - apart for monoatomic systems - and the sum of the volumes of the atoms should ideally approximate the volume of the cell. For additional details, we recommend to visit the web page https://www.vasp.at/wiki/index.php/RWIGS.[19] Hence, the atomic projection from a VASP calculation is only qualitative.

When projection from high pressure calculations is attempted, the standard procedure (*i.e.* projection via LORBIT = 11 in VASP, radii read from the POTCAR files) should be avoided, and the size of each atom should be carefully evaluated in order to cover the entire volume of the cell and avoid misleading projections. We propose a test that illustrates the importance of this statement. In Figure S1 we show the atomic projections perfomed with Crystal17, while in Figure S2 the one perfomed with VASP varying the size of the Wigner-Seitz radius. In one case, we have used one of the standard projections implemented in VASP (LORBIT = 11) using the 9 valence electron POTCAR of sodium (Figure S2a). The reader can notice that, according to this projection, Na appears to be an s and p valence element, as proposed by others in past theoretical studies on Na-hP4.[6,20] On the other hand, optimizing the radius of the sphere used for the sodium atom based on Bader's analysis (LORBIT = 1, RWIGS = 1.1), as shown in Figure S2b, a d-state for the valence electrons arises, as shown also by other past works.[3,21] Despite the fact that this radius is slightly longer than the shorter Na-Na distance in Na-hP4 measured at 190 GPa (see Figure S6), the covered volume is only 70.7% of the total volume, since the space in the cavities is not included. Increasing the radius to 1.24 Å covers the entire space (Figure S2c), but the projection changes again. However, with this distance the atomic spheres are inevitably overlapping. We have compared also this last projection with another one where 7 valence electrons of sodium were treated explicitly (Figure S2d). In this last case, very little difference can be seen when comparing with the 9 valence electron projection.

Finally, a few differences and similarities between the projections performed with Crystal17 and VASP can be pointed out. First, (with the exception of the standard projection, Figure S2a) both codes find that the s-p-d ratio (Figure S1-S2), is clearly dominated by pd-type states. However, d-states are even more prominent in the orbital based code. Moreover, the projection performed with Crystal17 seems to discriminate between the two symmetry inequivalent sodium atoms in the unit cell. In fact, their projected density of states are different (Figure S1). Meanwhile, VASP generates almost the same pDOS profile for both sodium atoms, despite the fact that they have two different chemical environments. The pDOS produced by Crystal17 gives us additional information: the valence electron DOS is primarily formed by pd-electrons of Na(1) (See Figure S1), suggesting a strong pd-hybridization. However, the overall shape of the total density of states calculated with Crystal17 and VASP (Figure S3), are almost identical, proving that the localized orbital basis is able to represent the electronic structure of this electride phase.

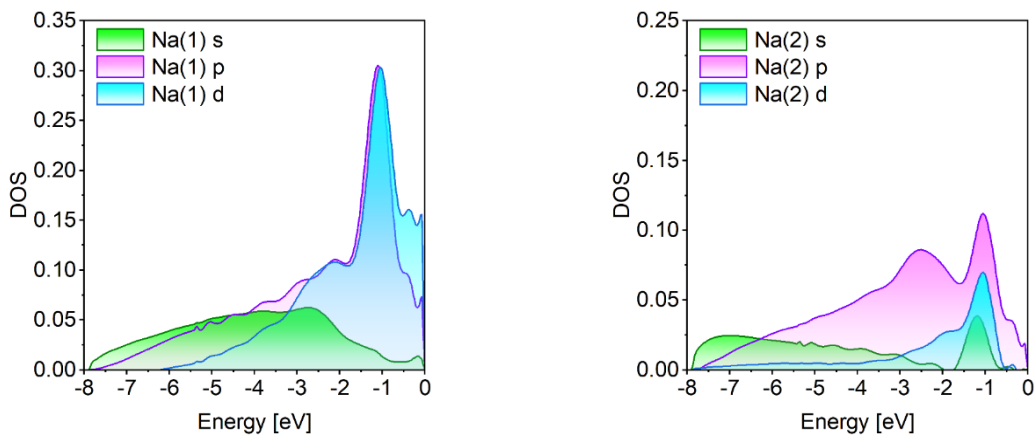

**Figure S1.** Projected density of states (pDOS) per atom of Na(1) and Na(2) in Na-hP4, calculated with the Crystal17 program package using the experimental geometry reported by Ma at 190 GPa,[3] with the HSE06 functional and the aforementioned *Na* basis (denoted as HSE06/*Na*). The zero of energy is at the top of the valence band.

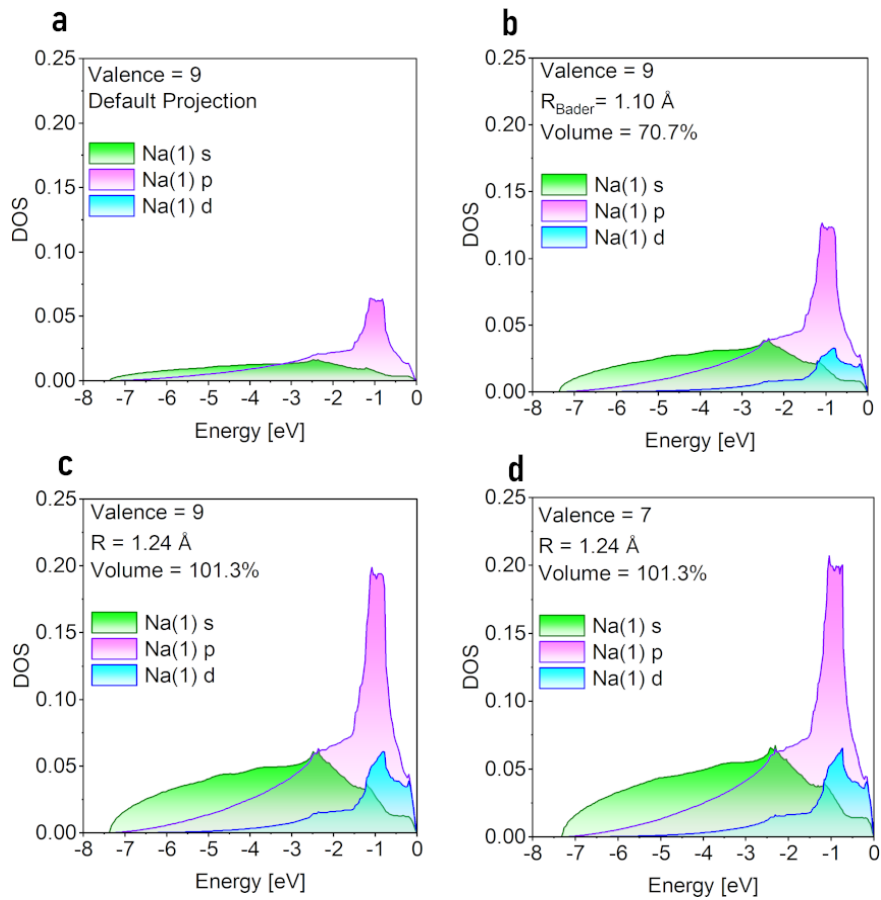

**Figure S2.** Projected DOS per atom of Na(1), which turns out to be the same as the pDOS calculated for Na(2), in Na-hP4 calculated with VASP, using the PBE[12,13] exchange-correlation functional (cut-off = 2000 eV) and explicitly treating 9 (**a,b,c**) or 7 (**d**) valence electrons of sodium at 190 GPa. We tested the (**a**) default Wigner-Seitz atomic radius varying it to values of (**b**) 1.10 Å, and (**c,d**) 1.24 Å. The percentage of the total unit cell volume covered by the spheres around each atom used by the projection is also reported (ideally it should be equal to 100%). The zero of energy is at top of the valence band.

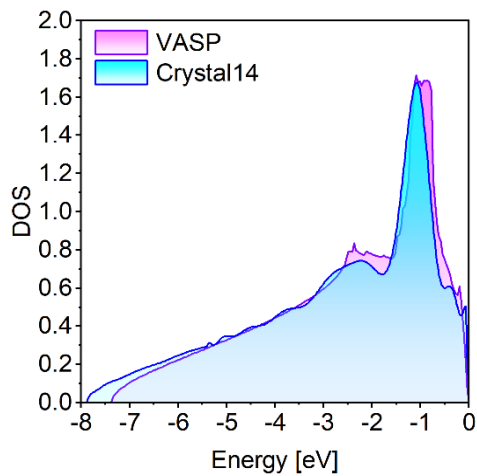

**Figure S3.** Comparison between the total DOS per unit cell calculated with VASP (HSE06 with 9 valence electrons) and Crystal17 (HSE06/*Na*).

### 3.0 Analysis of the importance of d-orbitals in Na-hP4

Codes that employ atomic orbital basis sets allow us to systematically study the effect of a specific type of orbital on the electronic properties. From the projected density of states shown above, it appears that d-orbitals are pivotal for the formation of the electride state in Na-hP4, as suggested also in past works.[3,4,21] However, this might not be the only factor important for electride formation. In fact, sodium might not even undergo an s-to-(p)d electronic transition, and some of its electrons might be expelled into ISQs centered in the cavities, forming electrides.[20] To test these possibilities, we analyzed the band-gaps and the properties of the electron density, through topological analysis and calculation of the electron localization functions (ELFs), using our optimized basis set (*Na*) and other possible basis-sets, *i.e.*, *Na-d*, *Na-pd*, *Na-d+1s*, *Na-pd+1s*, *Na-d+3s* and *Na-pd+3s*. To ensure an unbiased interpretation, when it was possible, we have compared these results with the same properties calculated with VASP by us, and others in previous works (*e.g.*, Ma *et al.* calculated a band gap of 1.3 eV at 200 GPa with the *GW* method).[3]

**3.1 Electron Localization Function and Band-gap**
In Figure S4, we provide a contour plot of the ELF along the (110) plane, calculated using Crystal17 with the HSE06 functional, for six basis-sets, together with the associated band-gaps (Table S1). The optimized basis-set, *Na*, shows a region of electron concentration in the interstitial site, together with a band-gap of 1.25 eV, which is close to the one calculated by Ma *et al.* with *GW* at a similar pressure. For comparison, the band-gaps calculated by VASP with the same functional (Table S1), are 1.23 eV, 1.24 eV and 1.25 eV, respectively, using PAWs that explicitly consider 9, 7 and 1 valence electrons. We can already note that the number of explicit electrons does not affect the band-gap of Na-hP4. Based on these results, we can also see that the optimized basis-set, *Na*, performs quite well in describing the electronic structure of Na-hP4.

Then, what would happen if we remove the valence 3d-orbitals only? This is the case of the *Na-d* basis, which can describe Na-hP4 using only 3s and 3p orbitals. Removal of the 3d-orbitals transforms Na-hP4 into a pure metal (Figure S4b and Table S1). The ELFs associated with the localized interstitial electron density disappear almost completely and the band-gap closes. Removing also the 3p orbitals (*Na-pd*), forcing the Na $3s^1$ configuration, the ELFs between Na atoms are completely removed (not reported in Figure S4). Obviously these cannot be good descriptions for Na-hP4, since we do expect a finite band-gap.[3,21] These results are in-line with the VASP calculations performed by us and others, where the 3d-orbitals clearly do play some role (Figure S2).

Further, according to the ISQ theory, the electrons in the cavities should be treated as free electrons at their ground state, and so, fully described by orbitals with low *n* and low *l* quantum numbers (1s, and perhaps 2s), while sodium should remain an s element.[5,20] This is another interesting possibility, which we checked by adding extra s orbitals to the cavities, while treating sodium with the *Na-pd* and *Na-d* basis-set. Following the ISQ theory, if we force sodium to retain a $3s^1$ electronic configuration, but allow the electron to move into the ISQ site by including extra s-functions in the interstitial sites (*Na-pd+1s* and *Na-pd+3s*) the first thing we notice is that the electron density associated with the ISQs reappears in the ELF plots! But they do not look like those in the plots calculated with the *Na* basis. Moreover, the calculated band-gaps are totally different (2.50 and 2.92 eV, Table S1), bearing a stronger resemblance to a large band gap insulator than to a semiconductor. Moreover, the ELFs become extremely localized (Figure S4c-d). Based on these results, it appears that thinking of the electride as being comprised of electrons localized in the cavities (computationally modelled via placing s-like orbitals centered on a fictitious ISQ coupled with the *Na-pd+1s* or *Na-pd+3s* basis-sets) is not correct, since the associated computational model does not provide the correct description for the electronic structure of Na-hP4.

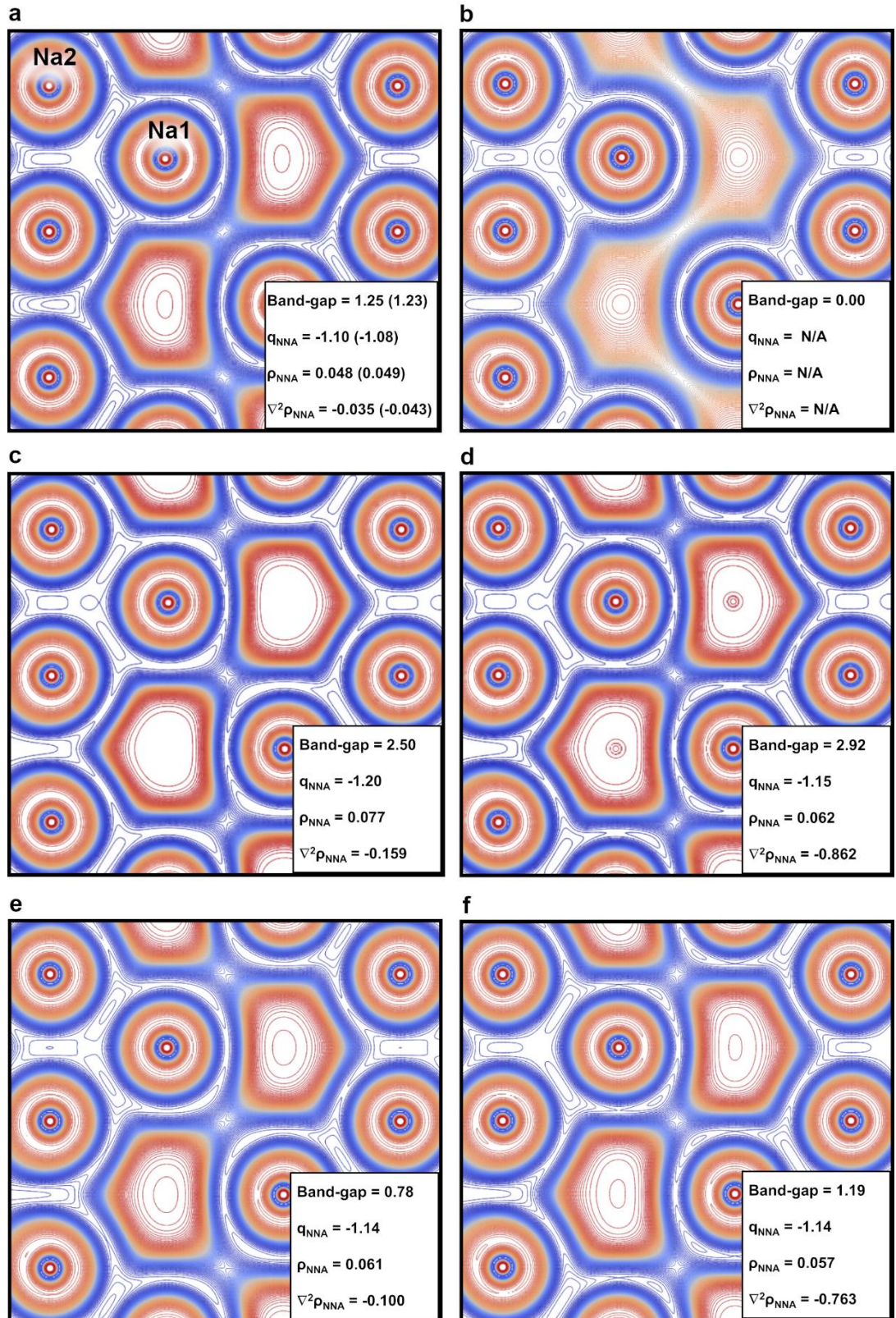

**Figure S4.** 2-D ELF isovalues along the (110) plane of Na-hP4 at 190 GPa calculated with different basis-sets: a) *Na*; b) *Na-d* (in *Na-pd*, the ELFs in the cavities disappear completely); c) *Na-pd+1s*; d) *Na-pd+3s;* e) *Na-d+1s*; f) *Na-d+3s*. The color scale goes from 0 (blue) to 1 (red). Each panel also lists the band-gap calculated with the HSE06 functional, the charge integrated in the NNA basin ($q_{NNA}$, [e]), and the values of the electron density ($\rho_{NNA}$, [e/bohr$^3$]) and Laplacian of the electron density ($\nabla^2\rho_{NNA}$, [e/bohr$^5$]) at the NNA position. Results from VASP code are in parenthesis in panel a).

In addition, we expanded even further these basis-sets including additional 2p orbitals in the ISQs (not reported in Figure S4), but these settings resulted again in large band-gap systems with unphysical double-concentration loci in the ELF plots.

The results obtained with the *Na-d* basis-set showed that the p-orbitals alone are not enough to describe the electronic properties of Na-hP4. However, if we allow sodium to both polarize with 3p functions and to transfer electrons to the cavities using the s-functions

(*Na-d+1s* and *Na-d+3s*), could we retrieve different results? With *Na-d+1s*, it was still not possible to calculate the correct band-gap, which is too small (0.78 eV, Table S1). However, providing greater flexibility to the basis-functions within the ISQ by adding two more s functions (*Na-d+3s*) helps to increase the band-gap to 1.19 eV, very close to the expected value. The ELF contour plot calculated for *Na-d+1s* and *Na-d+3s* are not exactly like those computed using the *Na* basis, being slightly more contracted towards the center of the ISQ, and showing the region between ELFs as being more separated (Figure S4e-f).

Nonetheless, the Na-d+3s basis seems to work quite well, both in terms of the calculated band-gap and ELF contours. However, the reason why the Na-d+3s basis describes the electride state is not clear; it could be because it allows the valence electrons of the sodium atoms to hop to the s-orbitals placed on the ISQ site, or it could be because the Na 3p orbitals could overlap with those of the ISQ, and in so doing undergo an artificial anisotropic expansion. Since ELF is mostly a qualitative analysis, we need a more quantitative method to evaluate if *Na-d+3s* and *Na* are two equivalent models, and if the d-orbitals are really necessary to correctly describe the electronic properties of Na-hP4. These questions are answered below.

**Table S1.** Band gaps, in eV, calculated at different levels of theory (funcitonal + basis-set/POTCARs) with Crystal17 and VASP.

| Crystal17 Functional | Basis-set | Band-Gap [eV] [a] |
|---|---|---|
| HSE06 | *Na* | 1.25 |
| HSE06 | *Na-d* | 0.00 (metallic) |
| HSE06 | *Na-d+1s* | 0.78 |
| HSE06 | *Na-d+3s* | 1.19 |
| HSE06 | *Na-pd* | 0.00 (metallic) |
| HSE06 | *Na-pd+1s* | 2.50 |
| HSE06 | *Na-pd+3s* | 2.92 |

| VASP Functional | POTCAR (#valence electrons) | Band-Gap [eV] |
|---|---|---|
| HSE06 | 9 | 1.23 |
| HSE06 | 7 | 1.24 |
| HSE06 | 1 | 1.25 |

[a] Reference value calculated by Ma *et al.* with GW at *ca.* 200 GPa = 1.3 eV.[3]

**3.2 Electron Density Analysis and Integration of the Bader Basins in the Electride**
The ELF analysis itself is a good indicator to check whether an electride state is formed. However, as shown by Postils *et al.*,[22] it should not be used alone, especially for unknown compounds, but it could possibly be coupled with a topological analysis to check for the presence of local maxima, or other topological features, and the character of the localization (magnitude of $\nabla^2\rho$).

Topological analysis of the electron density inside the cavities of Na-hP4, performed with Crystal17 and VASP (Table S2), yields the values of the electron density at the non-nuclear attractor positions (NNA), $\rho_{NNA}$, the values of the Laplacian at the same points, $\nabla^2\rho_{NNA}$, and the integrated Bader charges of the NNA basins. $\rho_{NNA}$ measures the value of the electron density at the NNA position, while $\nabla^2\rho_{NNA}$ quantifies how much the charge is concentrated/depleted at that point ($\nabla^2\rho_{NNA} < 0$ means a concentration of charge).

Our results show that HSE06/*Na* reproduces almost identically the results obtained with VASP, which does not depend much upon the number of valence electrons treated explicitly (Table S2). However, all the other models, which include extra s-functions in the interstitial sites and exclude p- and d-orbitals, including *Na-d+3s*, force the electron density in the cavities to be too localized (high absolute values of $\rho_{NNA}$ and $\nabla^2\rho_{NNA}$) and are very different from the results obtained with VASP. Therefore, d-orbitals are truly pivotal for the correct description of the localized electron in Na-hP4.

From both codes, the integration of Bader's basins revealed another interesting aspect: Each cavity hosts, on average, only one electron, not a pair. Bader charge is known to often overestimate the atomic charge of an atom, especially for covalent compounds.[23] However, it usually works very well for ionic systems.[24] At a first glance, this result might seem to be in disagreement with the ELF, which is often erroneously interpreted as a sole measurement of the localization of pairs of electrons.[25,26] The ELF is formally defined as a measure of the localization of Fermi-holes, or also as an estimate of the Pauli repulsion;[25,26] in other words, everything that is not a pair of electrons with the same spin.[27] This means that a single, isolated, electron can also generate a high value of the ELF. In fact, the ELF asymptotically tends to a value of 1 in regions of space dominated by a single orbital occupied by an unpaired electron.[25,26] To illustrate this, we show a contour plot of the ELF calculated for the hydrogen atom and for the hydrogen molecule in a triplet state (Figure S5). Therefore, the integration of one electron in the Bader basin is not in disagreement with the calculated ELF plots for Na-hP4. However, we must stress that the Bader basins possess a very clear physical meaning,[14] and the volumes produced by the ELF analysis[28] are not analogous quantities.

Finally, we want to highlight that explicitly treating 1, 7 or 9 valence electrons in the VASP calculations did not affect our conclusions, yielding almost identical values for the topological analysis, charges and band-gaps. This agrees with the discussion of the COOP and COHP analyses in the main text (See Table S3), where it is pointed out that the size of the core cannot affect the electrons in the cavities, since they are generated mainly by d- and p-orbitals, which are orthogonal to the core.

**Table S2.** Values of the electron density (ρ), Laplacian of the electron density ($\nabla^2\rho$), and integrated charge (q), in atomic units ([e/bohr$^3$], [e/bohr$^5$] and [e]), calcualted at the NNA (non-nuclear attractor) position using different levels of theory.

| Crystal17 Functional | Basis-set | $\rho_{NNA}$ | $\nabla^2\rho_{NNA}$ | $q_{NNA}$ |
|---|---|---|---|---|
| HSE06 | *Na* | 0.048 | -0.035 | -1.10 |
| HSE06 | *Na-d+1s* | 0.061 | -0.100 | -1.14 |
| HSE06 | *Na-d+3s* | 0.057 | -0.763 | -1.14 |
| HSE06 | *Na-pd+1s* | 0.077 | -0.159 | -1.20 |
| HSE06 | *Na-pd+3s* | 0.062 | -0.862 | -1.15 |

| VASP Functional | POTCAR (#valence electrons) | $\rho_{NNA}$ | $\nabla^2\rho_{NNA}$ | $q_{NNA}$ |
|---|---|---|---|---|
| HSE06 | 9 | 0.049 | -0.043 | -1.08 |
| HSE06 | 7 | 0.049 | -0.042 | -1.07 |
| HSE06 | 1 | 0.051 | -0.047 | -1.02 |

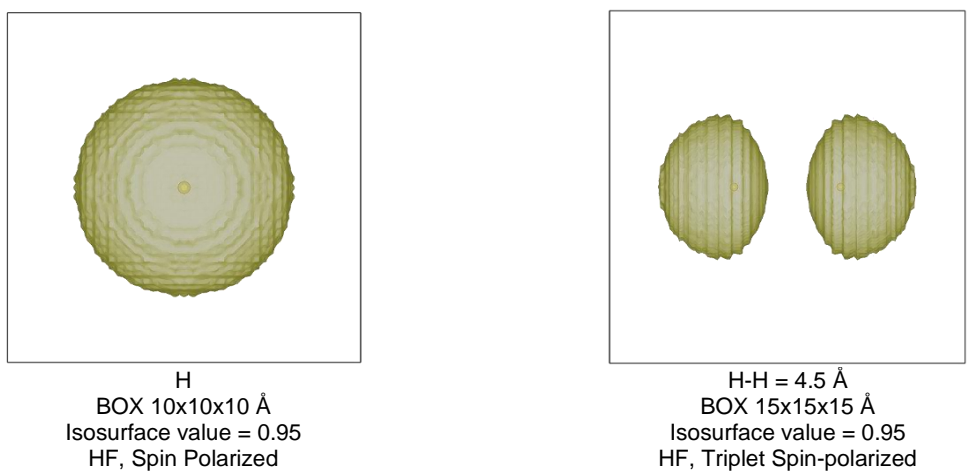

H
BOX 10x10x10 Å
Isosurface value = 0.95
HF, Spin Polarized

H-H = 4.5 Å
BOX 15x15x15 Å
Isosurface value = 0.95
HF, Triplet Spin-polarized

**Figure S5.** ELF isosurfaces of atomic hydrogen (1 electron) and hydrogen molecule in a triplet electronic state (2 electrons with the same spin), computed with VASP with Hartree-Fock (HF) in a spin polarized calculation and a cutoff of 500 eV.

**Table S3.** Intra-atomic crystal orbital overlap population (COOP) and negative of the crystal orbital Hamilton population (-ICOHP) integrated to the Fermi level (in eV/bond, where *bond* refers to orbital pairs within the same atom), calculated between intra-atomic orbitals for Na1 and Na2 in Na-hP4 with Crystal17 at the HSE06/*Na* level of theory.

| Orbitals | ICOOP(Na1) [a] | -ICOHP(Na1) [b] | ICOOP(Na2) [a] | -ICOHP(Na2) [b] |
|---|---|---|---|---|
| 3d ⋯ (1s+2s) | 0.0 | 0.2 | 0.0 | 0.0 |
| 3p ⋯ (1s+2s) | 0.0 | 0.1 | 0.0 | 0.0 |
| 3s ⋯ (1s+2s) | -4.8 | -178.4 | -4.8 | -179.5 |

[a] ICOOP > 0 means bonding, ICOOP < 0 means anti-bonding, and ICOOP = 0 means non-bonding. [b] The -ICOHP refers to the negative of the integrated value of the COHP over all the occupied states. This value is often shown because it refers to a stabilizing interaction when it is positive, and destabilizing when negative.

**4.0 The Topological Connectivity between Charge Densities**

Let us now present an analysis of the chemical bonds in Na-hP4 performed on the topology of the electron density using the QTAIM (See Figure S6 and Table S2 and S4). QTAIM is the cornerstone method to study chemical bonding in real space, both from theory and experiments.[26,29] One of the widely used ways to check whether a bond is present, together with its strength, is the analysis of the bond critical points (bcp) in the electron density. It has been well established that, unlike the name says, bond critical points do not always guarantee the presence of a chemical interaction.[30] However, they are still useful to investigate the shape of the electron density between atoms.

Our QTAIM analysis on Na-hP4 yielded several similarities with the studies performed by Gatti and Bader on alkali metal clusters.[31,32] Particularly, both our results and theirs showed that sodium atoms and NNAs are topologically connected. This last point can be seen in a plot of the Bader basins associated with the NNAs (or ISQs) together with the contour isoline, which shows that they are not separated by those of the sodium atoms (Figure S7). Therefore, from a topological point of view, the charge concentrations in the cavities do not seem to be isolated objects, but show some degree of delocalization.

**Table S4.** Values of the electron density ($\rho_b$) and Laplacian of the electron density ($\nabla^2\rho_b$), in atomic units, calculated at the bond critical points found by the topological analysis of the Na-hP4 electron density. Calculations were performed with VASP at the HSE06 level of theory with 9 valence electrons, and with Crystal17 at the HSE06 level of theory with 11 valence electrons in parenthesis (*Na* basis-set).

| Connection | d [Å] | $\rho_b$ [e/bohr$^3$] | $\nabla^2\rho_b$ [e/bohr$^5$] |
|---|---|---|---|
| Na(1)–Na(2) | 1.995 | 0.03 (0.03) | 0.26 (0.26) |
| Na(2)–Na(2) | 2.135 | 0.02 (0.02) | 0.16 (0.16) |
| Na(1)–NNA | 1.686 | 0.04 (0.03) | 0.07 (0.04) |
| Na(1)–NNA | 2.135 | 0.03 (0.03) | 0.08 (0.05) |

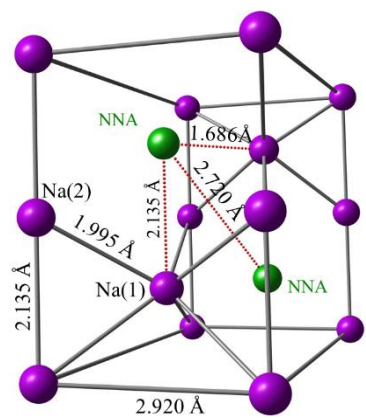

**Figure S6.** Sodium atomic position (purple spheres) in the Na-hP4 unit cell, reported together with the non-nuclear attractor (NNA) positions (green spheres) and the respective inter-atomic and inter-NNA (or inter-ISQ) distances at 190 GPa.

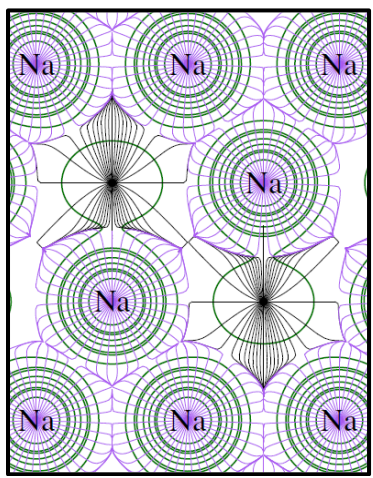

**Figure S7.** Isolines (in green) and gradient vectors (purple for sodium atoms and black for the NNAs) of the electron density of Na-hP4, calculated with VASP at the HSE06 level of theory and explicitly treating 9 valence electrons at 190 GPa.

**5.0 TB-LMTO and NMTO Calculations**
TB-LMTO calculations,[33] using the Na-hP4 structural parameters at 320 GPa, were performed using the von Barth-Hedin local exchange correlation potential.[34] Scalar relativistic effects were included. The basis set consisted of spd LMTOs on the Na atoms and sp LMTOs on the empty spheres (positions of the ISQs). The calculations utilized a 12x12x6 mesh in the tetrahedron *k*-space integrations. The version of the *N*MTO program employed[35] is not self-consistent and requires the output of the self-consistent potential from the TB-LMTO program. Figure S8a shows that the occupied bands obtained with the full *N*MTO basis set (red), using the specified energy mesh, yielded bands that were in good agreement with those calculated with VASP-PBE with a 9 electron POTCAR (black).

The downfolded *N*MTO band structures were then compared with the bands computed employing the full *N*MTO basis set; not with those obtained using the TB-LMTO program. Figure S8b shows the band structure calculated with $p_x$, $p_y$, $p_z$ orbitals placed on every Na atom, along with an s orbital placed on every empty sphere (black). This downfolded band structure agrees well with the one obtained with the full *N*MTO basis (red), and is able to describe *all* of the occupied bands. This includes the Na p-core, found in the range of -35 to -25 eV, as well as the valence band. To the right of the overlaid band structures we plot the Wannier-like function (WF) that is centered on the empty-sphere, and one of the resulting WFs on Na(1), noting that its shape and nodal structure look like a p orbital centered on Na with p-like orthogonalization tails on the nearest neighbor Na atoms. Finally, in Figure S8c we illustrate the band structure computed by only placing an s orbital on every empty sphere and chosing an energy mesh that spanned the valence band (black). In this calculation, all of the partial waves on the Na atoms were downfolded and the s like partial waves on the empty spheres were used in the basis. The resulting band structure is identical, on the scale of the figure, with the valence band obtained using the full NMTO basis. The resulting WF is plotted, and as described in the text, it could be formed from a constructive overlap of pd-hybrid orbitals of the 11 Na atoms that surround the empty sphere. Qualitatively, this WF is similar to the one obtained using the full occupied basis (Figure S8b).

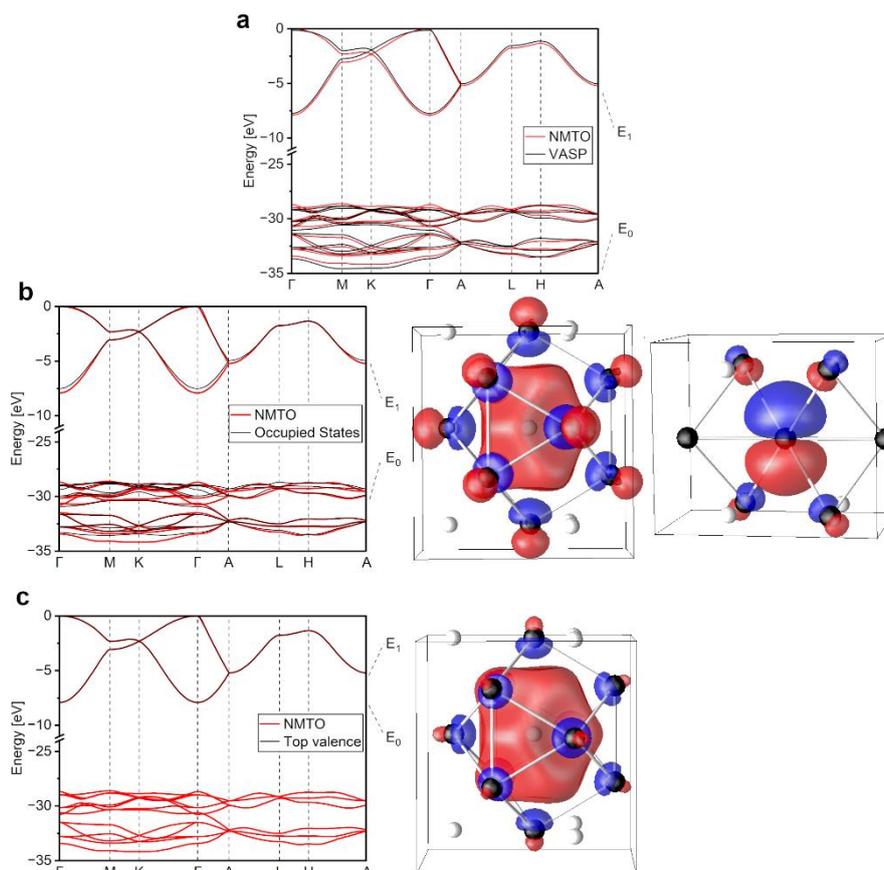

**Figure S8.** a) Electronic band structure calculated with VASP (PBE functional, 9 electron POTCAR) at 320 GPa is shown in black. In red we illustrate the NMTO band structure obtained using a full spd basis set on all of the Na atoms, and an sp basis on all of the empty spheres, along with two energy mesh points shown. The third energy mesh point (not shown) used for this full NMTO basis was $E_2 = 8.84$ eV. (b) The full NMTO band structure (black) is overlaid with the band structure obtained with a $p_x$, $p_y$ and $p_z$ orbital on each of the Na atoms in the unit cell, along with an s orbital on the empty sphere, and the energy mesh shown. Isosurfaces of the NMTO centered on the empty sphere and one of the Na p-like functions are illustrated. (c) The full NMTO band structure (black) is overlaid with the band structure obtained with an s orbital on the empty sphere, illustrated on the right, and the given energy mesh. Sodium atoms are black, empty spheres are white.

## 6.0 Thermodynamics

Comparison of thermodynamic terms, $\Delta H = \Delta E + \Delta PV$, where $H$ is the enthalpy, $E$ is the internal energy, $P$ is the pressure and $V$ is the volume of the Na-hP4 phase relative to the Na-bcc phase as a function of pressure is shown in Figure S9. The main contribution stabilizing Na-hP4 under pressure is the $PV$ term. In fact, despite the existence of the cavities in the hP4 phase, its density is 6% higher than that of bcc (4.819 g/cc for Na-hP4, vs. 4.555 g/cc for Na-bcc at 200 GPa). The formation of multicenter bonds, which leads to a concomitant localization of the electron density in the cavities, allows the sodium atoms in Na-hP4 to be closer to each other (Na-Na = 1.997 Å at 200 GPa), as compared to a bcc geometry, where the electron density of the bonds is concentrated between pairs of atoms, producing longer interatomic distances (Na-Na = 2.216 Å at 200 GPa).

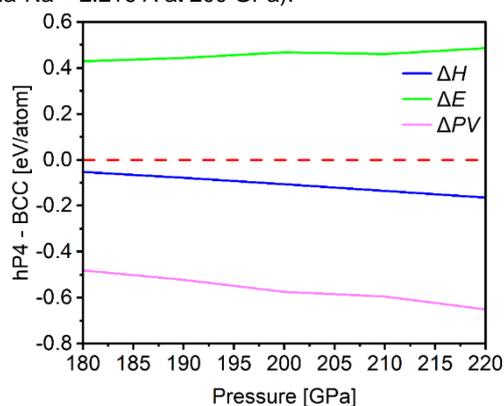

**Figure S9.** Thermodynamic terms $\Delta H$, $\Delta E$ and $\Delta PV$ as a function of pressure, associated with the transformation of the Na bcc phase into the hP4 phase.

## 7.0 General Aspects of the Effects of Orbital Hybridization

The hybridization of valence orbitals is crucial to provide the correct directionality to chemical bonds, allowing extended coordination numbers and multi-center sharing of electrons. Therefore, we tested the importance of orbital hybridization for an experimentally observed K-hP4 phase at 25 GPa[36] and an hypothetical Li-hP4 phase at 200 GPa. The projected densities of states plots (pDOS) shown below illustrate that the d-orbitals are the main contributors to the valence state in K-hP4, supporting our thesis for their importance in the generation and stabilization of the hP4 phase. There is no estimation for a range of stability of Li-hP4, and therefore, the orbital ordering we are presenting might by biased by our choice of pressure. However, we note that typically as one descends a column in the periodic table, similar phase transitions occur but at lower pressures, motivating our choice of 200 GPa for Li. The calculation of the pDOS in Li-hP4 shows that the valence states are almost completely given by the p-orbitals, with a strong hybridization with s-orbitals around -3 eV below the Fermi level. Therefore, also in Li-hP4 the electron density in the cavity necessities the hybridization with orbitals having higher angular momenta, which can provide the right directionality. In both cases, the plots of the ELF support the localization of charge in the same interstitial region and electride formation, but we note that whereas the Li phase is dynamically unstable, the K phase is computed to be a local minimum, in agreement with the experimental observations.[36] In both phases, we observe the localization of charge in the interstitials via the calculated ELF loci, however we note that they are generated by different orbitals.

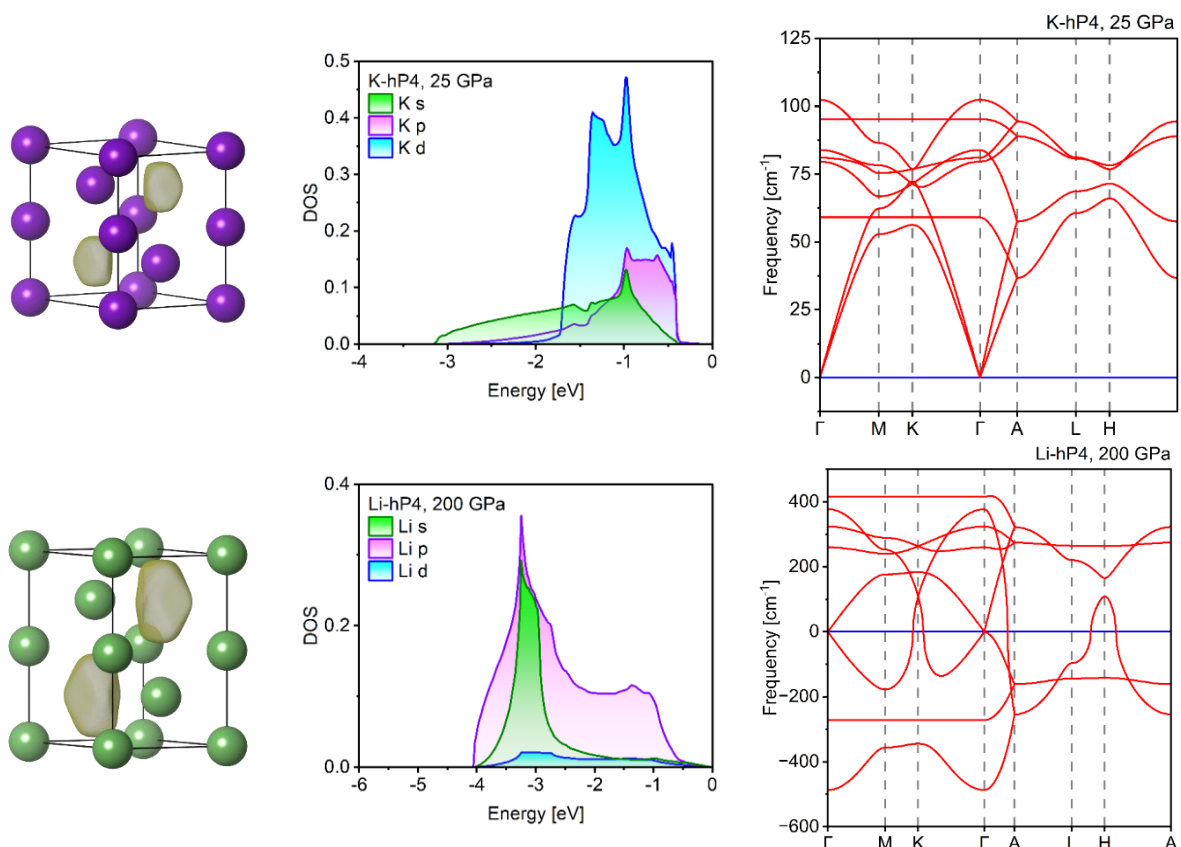

**Figure S10.** ELF isosurfaces with isovalue 0.8 and projected density of states of the proposed K-hP4 phase at 25 GPa[36], and of a hypothetical Li-hP4 phase at 200 GPa. Phonon calculations show that the former is dynamically stable, and the latter is not at the given pressures.